\def\papertitle{A Hierarchical Deep Learning Approach for Minority Instrument Detection}
\def\paperauthorA{Dylan Sechet}
\def\paperauthorB{Francesca Bugiotti}
\def\paperauthorC{Edouard d'H\'erouville}
\def\paperauthorD{Filip Langiewicz}
\def\paperauthorE{Matthieu Kowalski}

\documentclass[twoside,a4paper]{article}
\usepackage{booktabs}
\usepackage{adjustbox}
\usepackage{dsfont}
\usepackage{etoolbox}
\usepackage{dafx_24}
\usepackage{tikz}
\usepackage{amsmath,amssymb,amsfonts,amsthm}
\usepackage{euscript}
\usepackage{svg}
\usepackage{graphicx}
\usepackage[T1]{fontenc}
\usepackage[utf8]{inputenc}
\usepackage{ifpdf}
\usepackage[english]{babel}
\usepackage{caption}
\usepackage{subfig} 
\usepackage{color}
\definecolor{linkcolor}{RGB}{83,83,182} 

\input glyphtounicode
\ninept

\newcounter{numauth}
\newcounter{listcnt}
\newcommand\authcnt[1]{\ifdefined#1 \stepcounter{numauth} \fi}

\newcommand\addauth[1]{
\ifdefined#1 
\stepcounter{listcnt}
\ifnum \value{listcnt}<\value{numauth}
\appto\authorslist{, #1}
\else
\appto\authorslist{~and~#1}
\fi
\fi}
\authcnt{\paperauthorB}
\authcnt{\paperauthorC}
\def\authorslist{\paperauthorA}
\addauth{\paperauthorB}
\addauth{\paperauthorC}
\addauth{\paperauthorD}
\addauth{\paperauthorE}

\usepackage{times}

\newif\ifpdf
\ifx\pdfoutput\relax
\else
   \ifcase\pdfoutput
      \pdffalse
   \else
      \pdftrue
\fi

\ifpdf 
  \usepackage[pdftex,
    pdftitle={\papertitle},
    pdfauthor={\authorslist},
    pdfsubject={Proceedings of the 27th International Conference on Digital Audio Effects (DAFx24)},
    colorlinks=false, 
    bookmarksnumbered, 
    pdfstartview=XYZ 
  ]{hyperref}
  \hypersetup{
    colorlinks=true,
    citecolor=linkcolor,
    linkcolor=linkcolor
  }
    \usepackage[nameinlink,capitalize]{cleveref}

\else 
  \usepackage[dvips]{epsfig,graphicx}
  \usepackage[dvips,
    pdftitle={\papertitle},
    pdfauthor={\authorslist},
    pdfsubject={Proceedings of the 27th International Conference on Digital Audio Effects (DAFx24)},
    colorlinks=false, 
    bookmarksnumbered, 
    pdfstartview=XYZ 
  ]{hyperref}
\fi
\usepackage[hypcap=true]{caption}
\title{\papertitle}

\twoaffiliations{
 \paperauthorA,  \paperauthorB\ and \paperauthorE \sthanks{We thank and acknowledge Agathe Gioan,
Xavier Jeunot, and
Aaron Broderick, for the productive discussions and the joint work that preceded this study.
}}
 {CentraleSupelec,  Inria, CNRS, LISN \\ Universit\'e Paris-Saclay, \\ Paris-Saclay, France\\
{\tt dylan.sechet@student-cs.fr}
 {\tt francesca.bugiotti@centralesupelec.fr}
{\tt matthieu.kowalski@universite-paris-saclay.fr}
 }
{\paperauthorC \, and \paperauthorD}
{\href{https://linkaband.com/}{Linkaband} \\ Paris, France \\ 
{\tt forename@linkaband.com}
}

\usepackage{microtype}

\begin{document}
\ifpdf 
\else  
  \DeclareGraphicsExtensions{.eps}
\fi


\maketitle

\begin{abstract}
  Identifying instrument activities within audio excerpts is vital in music information retrieval, with significant implications for music cataloging and discovery. Prior deep learning endeavors in musical instrument recognition have predominantly emphasized instrument classes with ample data availability. Recent studies have demonstrated the applicability of hierarchical classification in detecting instrument activities in orchestral music, even with limited fine-grained annotations at the instrument level. Based on the  Hornbostel-Sachs classification, such a hierarchical classification system is evaluated using the MedleyDB dataset, renowned for its diversity and richness concerning various instruments and music genres. This work presents various strategies to integrate hierarchical structures into models and tests a new class of models for hierarchical music prediction. This study showcases more reliable coarse-level instrument detection by bridging the gap between detailed instrument identification and group-level recognition, paving the way for further advancements in this domain.
  %
\end{abstract}

\section{Introduction}\label{sec:intro}
The identification of instruments within an audio excerpt poses an enduring challenge in the field of Music Information Retrieval (MIR). This task is inherently intricate due to the poly-instrumental nature of real-world music, where the pitches of multiple instruments often intertwine. Furthermore, the task is complicated by substantial variations in timbre and performance style among instruments, further hindering recognition endeavors. Even trained musicians may encounter perceptual similarities among specific instruments, adding another layer of complexity to the recognition process.

Instrument identification bears significant implications across various domains, including music cataloging and discovery. It aids in tasks such as song retrieval~\cite{ferreira2016accuracy}, facilitates genre recognition systems~\cite{sturm2013classification}, and contributes to recommendation systems~\cite{song2012survey}. While the recognition of more common instruments benefits from abundant available data, challenges arise in genres such as orchestral or opera, as well as with rare or non-western instruments, for which data is much more scarce.

Hierarchical classification systems have been proposed to address the complexities of instrument recognition. These systems enable the prediction of instruments at various levels of specificity, demonstrating particular promise in handling imbalanced datasets and scenarios involving few-shot learning~\cite{garcia_leveraging_2021}. However, existing work in this domain has been confined to specific genres and a restricted set of instruments. This study assesses the scalability of hierarchical approaches on a more complex dataset, like MedleyDB, and proposes different formulations of the hierarchical problem.

\subsection{Instrument detection}
The field of instrument detection incorporates a diverse array of methodologies, spanning from signal processing techniques to contemporary deep learning approaches. Meanwhile, multi-label classification for audio signals has attracted considerable interest across various domains~\cite{reghunath_transformer-based_2022}.

In the 2000s, Marques et al.~\cite{marques_study_nodate} conducted instrument classification on brief $0.2s$ music excerpts utilizing Gaussian Mixture Models (GMM) and Support Vector Machines (SVM), with features extracted through Mel-Frequency Cepstral Coefficients (MFCC). In a similar vein, Essid et al.~\cite{essid_musical_nodate} showcased the advantages of GMMs over SVMs by employing MFCC features preprocessed with Principal Component Analysis (PCA), even in the context of longer mono-instrument samples.

More recently, Deep learning models, which have seen successful applications across various domains~\cite{sarker_deep_2021}, have demonstrated promise in mono-instrument detection as well. Initially designed for image recognition tasks, Convolutional Neural Networks~(CNNs) have been effectively repurposed to handle spectrogram-like features such as MFCCs or Constant-Q Transforms. Solanky et al. highlighted in~\cite{solanki_music_2019} the efficiency of an AlexNet-inspired CNN model in predominant instrument recognition, while in~\cite{avramidis_deep_2021}, Avramidis et al. introduced performance enhancements in instrument recognition by integrating recurrent components into CNN architectures.

Attention-based models have also emerged as an promising alternative to CNNs in the field of audio classification: transformers, initially introduced for text classification~\cite{vaswani_attention_2017}, have been adapted to image recognition tasks~\cite{dosovitskiy_image_2021}. Jamil et al.~\cite{jamil_distinguishing_2022} used a vision transformer architecture for audio classification to distinguish harmless from malicious drones. In the domain of Music Information Retrieval, Regunath et al.~\cite{reghunath_transformer-based_2022} were able to outperform a CNN architecture using a vision transformer for predominant instrument recognition in polyphonic settings.

\subsection{Hierarchical classification for audio}
Exploration of hierarchical structures for classifying audio segments has been a subject of prior research across diverse domains. For instance, hierarchical methods have been employed in classifying bird songs, with a class tree rooted in biological taxonomy~\cite{cramer_chirping_2020}. Notably, these methods operated on more extended audio excerpts than the frame-level analysis.

In the realm of Music Information Retrieval, Fu et al. introduced in~\cite{fu_hierarchical_2019} a hierarchical approach tailored for singing voice classification and transcription. On the other hand, Essid et al.~\cite{essid_musical_nodate} explored hierarchical classification using GMMs, focusing primarily on synthetic music extracts rather than actual recordings and not at the frame level.
In a recent study in 2023, Krause et al.~\cite{krause_hierarchical_2023} delved into hierarchical classification methods explicitly designed for orchestral and opera pieces. Their research demonstrated performance improvements, particularly in scenarios with limited fine-grained annotations.
Furthermore, Garcia et al.~\cite{garcia_leveraging_2021} investigated hierarchical classification for few-shot learning situations, aiming to enable model adaptation to unseen classes, contrasting with the study's utilization of predefined classes.

\subsection{Work on rare instrument detection}
In the domain of rare audio source detection, where annotated data is scarce or nonexistent, previous research efforts have aimed to address this challenging task.
Various strategies have been explored in the domain of few-shot learning, including hierarchical methodologies akin to those proposed by Garcia et al.~\cite{garcia_leveraging_2021}, along with approaches centered on continual learning~\cite{wang_few-shot_2021}. Continual learning techniques focus on easily incorporating new instruments into a model as additional data becomes available. Moreover, attempts have been made to capitalize on weakly annotated data, where instrument presence is identified but precise activation times are not specified, yielding only incremental enhancements~\cite{mukhedkar_polyphonic_nodate}.

The utilization of pre-training strategies has emerged as a pivotal area of interest. In 2023, Zong et al.~\cite{zhong_exploring_2023} employed isolated notes for pre-training before transitioning to training on polyphonic data, albeit with a specific emphasis on predominant instrument recognition exclusively. Another explored avenue involves synthetic data generation achieved through layering mono-instrumental excerpts with tempo and pitch shifting to produce realistic artificial multi-instrument tracks~\cite{kratimenos_augmentation_2021}.

Furthermore, model reprogramming has been proposed, involving training a smaller model to map inputs to the input space of a larger pre-trained model. This technique, akin to transfer learning, harnesses the generalization capabilities of the larger model to mitigate data imbalance, consequently significantly reducing training time requirements~\cite{chen_music_2022}.

\subsection{Contributions and outline of the paper}
In this study, we introduce several methodologies aimed at efficiently incorporating hierarchical instrument structures into our predictive models, and evaluate this novel class of models tailored for hierarchical music prediction. Importantly, our evaluations are conducted on the MedleyDB dataset. This dataset is renowned for its expansive and varied content, which allows us to overcome constraints related to particular music genres and a restricted instrument set.
As far as we know, this is the first work on polyphonic instrument recognition using the MedleyDB dataset, providing crucial baseline performances in this domain.

The paper is organized as follows. \cref{sec:dataset} offers an overview of the MedleyDB dataset used in our study while \cref{sec:model} delves into the neural network architecture selected for our research. In \cref{sec:training}, we explore the various training strategies implemented to address the hierarchical structures of instruments. The numerical results derived from our evaluations are presented in~\cref{sec:results}. Finally, the conclusions and insights are summarized in \cref{sec:ccl}.

\section{Hierarchical dataset}\label{sec:dataset}
This section focuses on the hierarchical dataset utilized in our study. The primary dataset of interest is MedleyDB, emphasizing its characteristics and composition. We then discuss the challenges of establishing a train/test split for MedleyDB to prevent overfitting and ensure a balanced instrument distribution within the sets. Finally, we introduce a labeling scheme incorporating instrument group hierarchies and utilize the Hornbostel-Sachs classification system to categorize instruments based on sound production methods, balancing granularity and computational efficiency.

\subsection{MedleyDB}
\begin{figure}[!ht]
  \centerline{\includegraphics[width=\linewidth]{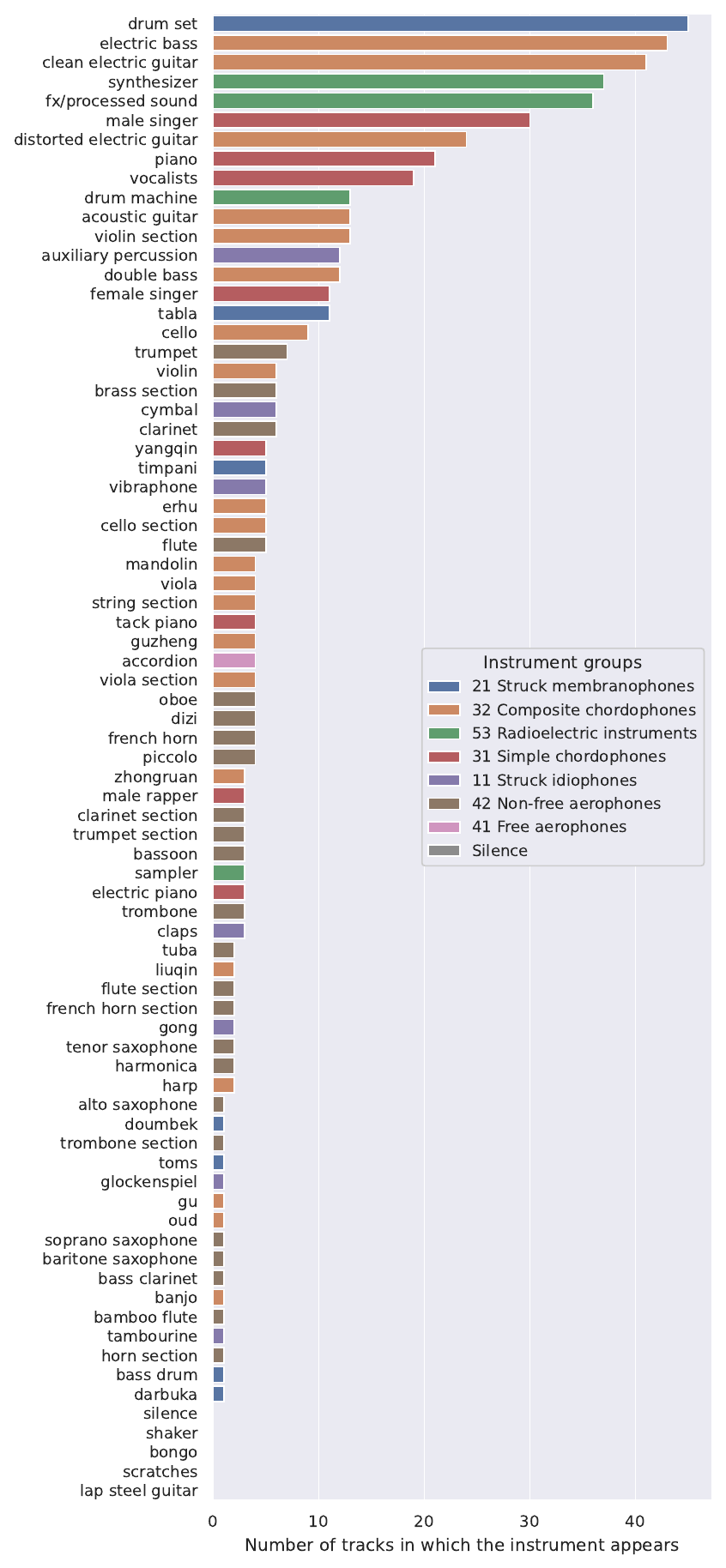}}
  \caption{\label{fig:medleydb}{\it Tail-heavy distribution of MedleyDB instrument occurrence. The numbers in the legend refer to the Hornobel-Sachs taxonomy.}}
\end{figure}

MedleyDB~\cite{bittner_medleydb_2014} is a dataset of $122$ annotated polyphonic recordings containing a large diversity of genres and instruments. It was curated primarily to support research on melody extraction by providing melody f0 annotations, but each track also contains precise instrument activations, making it usable for instrument recognition. The dataset is filtered to only include the $94$ tracks with no instrumental bleeding, in order to prevent erroneous instrument activation detections.
As seen in~\cref{fig:medleydb}, the distribution of instruments in MedleyDB is quite tail-heavy, featuring many instruments that appear in only a few of the tracks. This makes the resulting dataset extremely challenging. Indeed, it contains nearly as many instruments as tracks, which means the rarer instruments are usually showcased in a minimal context only.

We divide each track into non-overlapping frames of one-second duration. We consider an instrument active within a frame if it is active at any point during that duration. This method raises concerns regarding the potential misclassification of instruments if they are only briefly active at the beginning or end of a frame.  However, our analysis demonstrates that such occurrences are rare, with less than $0.26\%$ of frames containing an instrument active for less than $0.1$~second.

\subsection{Data train-test split}
We split the MedleyDB dataset into train and test recordings to train and evaluate our MIR system. To our knowledge, no standard train/test split has been established for MedleyDB in prior work. Establishing such a split is challenging: using extracts from the same song in both training and testing has been shown to lead to overfitting, but some instruments are very rare and only appear in a single recording. Ensuring a similar label distribution then becomes quite challenging, especially given that we are working with only $94$ tracks for $76$ instruments. Ultimately, we select $20$\% of the recordings to ensure a similar instrument distribution between both sets. Due to the inclusion of a $17$-minute long recording in the test set, we obtained a test set that is slightly larger than expected, with $14000$ excerpts in the training set and $5000$ in the test set.
We were unable to pick an alternate split to reduce the test set size, as alternatives resulted in strong instrument distribution shifts between the training and the test data. \textcolor{black}{This strong constraint on the dataset also made us unable to use $k$-fold validation.}
Special care is further taken to ensure the test set features various music genres. In the end, four instruments, each appearing in a single track, are present only at test time.

\subsection{Hierarchical classification}

We specify the labels further than a simple instrument name, adding labels per instrument group.
We therefore end up with two primary sets of labels: $\mathcal{I}$, indicating the instrument's name, and $\mathcal{G}$, containing labels for instrument groups.

The hierarchical classification system selected is referred to as Hornbostel-Sachs~\cite{von_hornbostel_systematik_1914}, organizing instruments according to their sound production method. This classification system is versatile and can effectively categorize a wide array of instruments from diverse cultural backgrounds. Its adaptability is particularly advantageous for datasets like ours, which are characterized by diverse instruments. Moreover, the Hornbostel-Sachs features up to five levels of depth, enabling us to configure the level of precision of the tree easily. For this study, we opted for a depth of two, balancing granularity and computational efficiency. \textcolor{black}{With this configuration, we split all instruments into $8$ different groups, shown in \cref{fig:medleydb}.}

However, this classification system has its drawbacks. Indeed, categorizing instruments based on their sound production method does not directly account for the output sound profile. This aspect poses challenges, especially when dealing with synthesized sounds. For instance, under this taxonomy, a drum machine would fall into a distinct class from a traditional drum despite producing similar sounds.
\textcolor{black}{We chose not to address these challenges specifically, as differentiating between synthetic and acoustic instruments may be required in certain contexts. The taxonomy can easily be adapted for tasks that do not require such a distinction.}

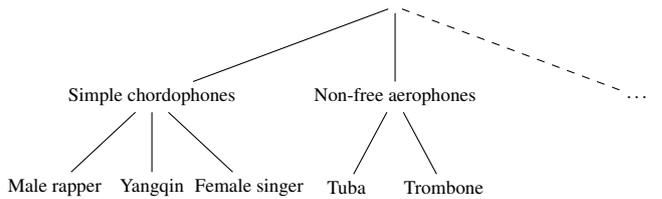
\begin{figure}[!ht]
  \centerline{

    \begin{tikzpicture}[level distance=1.5cm,
        level 1/.style={sibling distance=4cm},
        level 2/.style={sibling distance=1.6cm},
        scale=0.8, every node/.style={scale=0.8}
      ]
      \node {}
      child {
          node {Simple chordophones}
          child { node {Male rapper} }
          child { node {Yangqin} }
          child { node {Female singer} }
        }
      child {
          node {Non-free aerophones}
          child { node {Tuba} }
          child { node {Trombone} }
        }
      child[dashed] {node {\ldots}};
    \end{tikzpicture}
  }
  \caption{\label{fig:Hornbostel_sachs}{\it Partial representation of the Hornbostel-Sachs class tree}}
\end{figure}

\section{Model architecture}\label{sec:model}
In this section, we give details on the model's architecture used as the base brick of our hierarchical classification system.

Note that the model is not the primary focus of the paper, and alternative architectures (e.g., based on ResNets \cite{he_deep_2015}) could also be used here. We employ a convolutional network inspired by the VGG architecture \cite{simonyan_very_2015}, featuring a series of conv-conv-pool processing blocks. This architecture was chosen because VGGish models have shown good performance for MIR from spectral features for various downstream tasks \cite{grollmisch_analyzing_2021}, and require significantly less computing performance than transformer-based approaches. This series of processing blocks create feature maps of depth $64$, then $128$, and finally $256$ while aggregating context along the temporal and pitch dimensions and are followed by a standard classification head. Batch normalization is further used after each layer for regularization, and a leaky ReLU is used for activation. We finally apply dropout before fully connected layers in the classification head. The exact architecture is specified in \cref{tab:archi}.

The network takes MFCC of an audio extract as input and outputs a vector of $85$ values in $[0, 1]$ corresponding in activities of all classes in $\mathcal{I} \cup \mathcal{G}$. The MFCC input was chosen to consist of $1$~s of audio segments, computed using a hop-size of $1$s on recordings sampled at $22.5$~kHz, using $80$ bins.

\begin{table}[!ht]
  \centering
  \begin{tabular}{@{}ccc@{}}
    \toprule
    \textbf{Layer}      & \textbf{Output shape} & \textbf{Parameters}  \\ \midrule
    Input               & (1, 80, 22)           &                      \\ \midrule\midrule
    Conv2d              & (64, 80, 22)          & 640                  \\
    Batch normalization & (64, 80, 22)          & 128                  \\
    Conv2d              & (64, 80, 22)          & 36 928               \\
    Batch normalization & (64, 80, 22)          & 128                  \\
    MaxPool2d           & (64, 40, 11)          &                      \\ \midrule
    Conv2d              & (128, 40, 11)         & 73 856               \\
    Batch normalization & (128, 40, 11)         & 256                  \\
    Conv2d              & (128, 40, 11)         & 147 584              \\
    Batch normalization & (128, 40, 11)         & 256                  \\
    MaxPool2d           & (128, 20, 5)          &                      \\ \midrule
    Conv2d              & (256, 20, 5)          & 295 168              \\
    Batch normalization & (256, 20, 5)          & 512                  \\
    Conv2d              & (256, 20, 5)          & 590 080              \\
    Batch normalization & (256, 20, 5)          & 512                  \\
    MaxPool2d           & (256, 6, 1)           &                      \\ \midrule
    Conv2d              & (256, 1, 1)           & 393 472              \\
    Batch normalization & (256, 1, 1)           & 512                  \\
    Squeeze             & (256)                 & \multicolumn{1}{l}{} \\ \midrule
    Dropout             & (256)                 &                      \\
    Dense               & (256)                 & 65 792               \\
    Dropout             & (256)                 &                      \\
    Dense               & (128)                 & 32 896               \\
    Dropout             & (128)                 &                      \\
    Dense               & (85)                  & 10 965               \\ \midrule\midrule
    Output: Sigmoid     & (85)                  &                      \\ \bottomrule
  \end{tabular}
  \caption{\label{tab:archi}{\it Model architecture used for our classification system. Leaky ReLUs are used as the activation function.}}
\end{table}

\section{Model training}\label{sec:training}
We have tested four different approaches to model training in a hierarchical context, which we highlight here. We start by focusing on the impact of various loss functions, before introducing a new multi-model architecture\footnote{The code to train our model is publicly available on \href{https://github.com/Seon82/musedetect}{github}.}.

\subsection{Standard approach}
In our initial approach, we treat the labels from the combined set $\mathcal{I} \cup \mathcal{G}$ as a unified entity and train the model on these grouped labels. This method has the advantage of being relatively straightforward but completely disregards the inherent hierarchical structure within the data. Consequently, it may yield inconsistent predictions, as nothing prevents the model from mistakenly predicting a group label along with an instrument that doesn't belong to that group.

As a first approach, we train a model using a standard cross-entropy loss. To counterbalance the pronounced class imbalance within the dataset,  the loss is reweighted by inverse label frequency.
This standard loss reweighting technique forms a good baseline but remains extremely limited. Therefore, we also test a loss built specifically for imbalanced datasets, the focal loss $\mathcal{L}_{f}$. This loss, initially defined for object detection \cite{lin_focal_2018}, is defined as a slight variation on the cross-entropy loss:

\begin{equation}
  \mathcal{L}_{f}(\hat{y}, y) = - (1 - p_t(\hat{y}, y))\textcolor{black}{^2} \cdot \log(p_t(\hat{y}, y))
\end{equation}

with $p_t$ the predicted probability of the correct class.
This loss function has the advantage of dynamically giving more importance to misclassified samples during training. Indeed, for a sample classified correctly with high confidence, $1-p_t$ nears $0$, which causes the term to have little impact on the loss.

\begin{figure}[!ht]
  \centerline{\includegraphics[width=\linewidth,height=4cm]{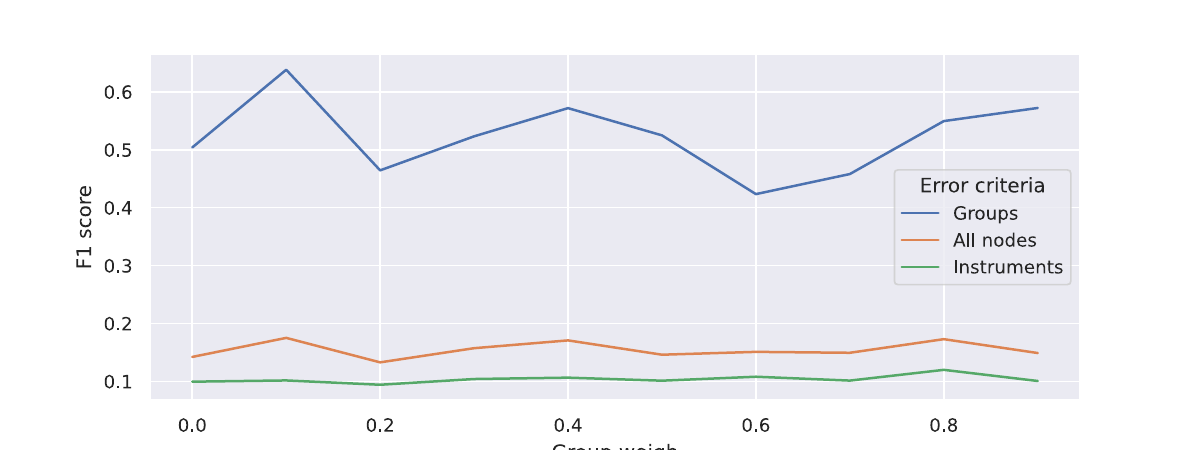}}
  \caption{\label{fig:alpha}{\it{F1-score depending on alpha value for} $\mathcal{L}_{weighted}$.}}
\end{figure}

This approach makes for a good baseline, but is unable to treat the labels in $\mathcal{I}$ and $\mathcal{G}$ differently. In a second approach, we attempt to apply a weight to each tree level in the loss function. For a given loss function $\mathcal{L}$, we define:

\begin{equation}
  \mathcal{L}_{weighted}(\hat{y}, y) = \mathds{1}_{\mathcal{I}}(y) \cdot \alpha \mathcal{L}(\hat{y}, y) + \mathds{1}_{\mathcal{G}}(y) \cdot (1-\alpha) \mathcal{L}(\hat{y}, y).
\end{equation}
We then run a grid search for different $\alpha$ values, with  $\mathcal{L}$  a cross-entropy loss. The results are presented in \cref{fig:alpha}. The maximal F1-score across all nodes is obtained for $\alpha = 0.1$, that is, putting much more emphasis on the group-level loss term. The curve presented in \cref{fig:alpha}, however, shows no clear trend.

Every model is trained for $30$ epochs using the Adam optimizer with a batch size of $32$ and a learning rate of $0.001$. The training was done on an Nvidia 1660Ti card, with each model taking around 30 minutes to train.

\subsection{Specialized models}
The previous approaches remained limited by treating labels from $\mathcal{I}$ and $\mathcal{G}$ as interchangeable, without any true accounting for the data's hierarchical structure. To overcome this problem and improve performance, we abandon the idea of a generalized model predicting groups and instruments in one pass and instead build a two-pass prediction system. To do so, we define a first model for group prediction, followed by specialized models for instrument prediction within each group. We train eight models using this approach: one group model trained with labels from $\mathcal{G}$ and seven specialized models, each predicting a subset of instruments from $\mathcal{I}$.

For simplicity, all models use the same VGG-like architecture, and all models are trained using the focal loss on the entirety of the dataset.
This model has a much greater capacity than the baseline models. However, artificially increasing the capacity of the baseline models (by adding two extra conv-conv-pool blocks) shows no significant performance increase, which allows us to suggest that any changes in performance are due to the change in architecture, not in capacity.
At inference time, the models are run in succession: the group-level model is run first, followed by each instrument-level model. This has a significant impact on inference speeds, making them eight times slower. The effect could likely be mitigated by implementing a gating structure, and only calling the instrument-level models if the group-level prediction is above a given threshold.

\section{Results and discussion}\label{sec:results}
Considering the pronounced data imbalance and the absence of prioritization between false positives and negatives within the application, we use the F1-score metric \cite{rijsbergen1979information} for evaluation:

\begin{equation}
  F1 = \frac{2 \times \text{precision} \times \text{recall}}{\text{precision} + \text{recall}\ }.
\end{equation}

As shown in \cref{tab:results}, the balanced cross-entropy performs similarly to the focal loss, with the latter having a slightly better performance for instrument prediction. This is unsurprising, given that the focal loss allows reweighting at a label granularity rather than simply for the tree levels. The fact that both performances are similar suggests that the focal loss' primary role is probably in rebalancing loss terms between group and instrument-level labels.

On the other hand, the weighted cross-entropy approach shows inferior performance and fails to learn instrument labels. Overall, we notice that performance for groups is significantly higher than for instruments across all models. This result is in accordance with our initial expectations, given that the reason for implementing groups was hopes for better performance in groups even when fine-grain instrument detection is unachievable.

\begin{table}[!ht]

  \centering
  \begin{adjustbox}{width=.5\textwidth}
    \begin{tabular}{cccc|ccc}
      \hline
      \textbf{}                & \multicolumn{3}{c|}{\textbf{Groups}} & \multicolumn{3}{c}{\textbf{Instruments}}                                                                  \\ \hline
                               & F1                                   & Precision                                & Recall        & F1             & Precision     & Recall        \\ \hline\hline
      Balanced cross-entropy   & 0.74                                 & \textbf{0.76}                            & 0.72          & 0.41           & \textbf{0.53} & 0.35          \\
      Focal loss               & 0.74                                 & \textbf{0.76}                            & 0.73          & 0.43           & 0.52          & 0.37          \\
      Weighted cross-entropy   & 0.64                                 & 0.51                                     & \textbf{0.86} & 0.17           & 0.52          & 0.06          \\
      Group-specialized models & \textbf{0.78}                        & \textbf{0.76}                            & 0.81          & \textbf{0.45 } & 0.50          & \textbf{0.40} \\ \hline
    \end{tabular}
  \end{adjustbox}
  \caption{\label{tab:results}{\it Performance of the different models. Averages are micro-averages, giving equal weight to each sample. The best method for each metric uses boldface.}}
\end{table}

\begin{figure}[!ht]
  \centerline{\includegraphics[width=.9\linewidth]{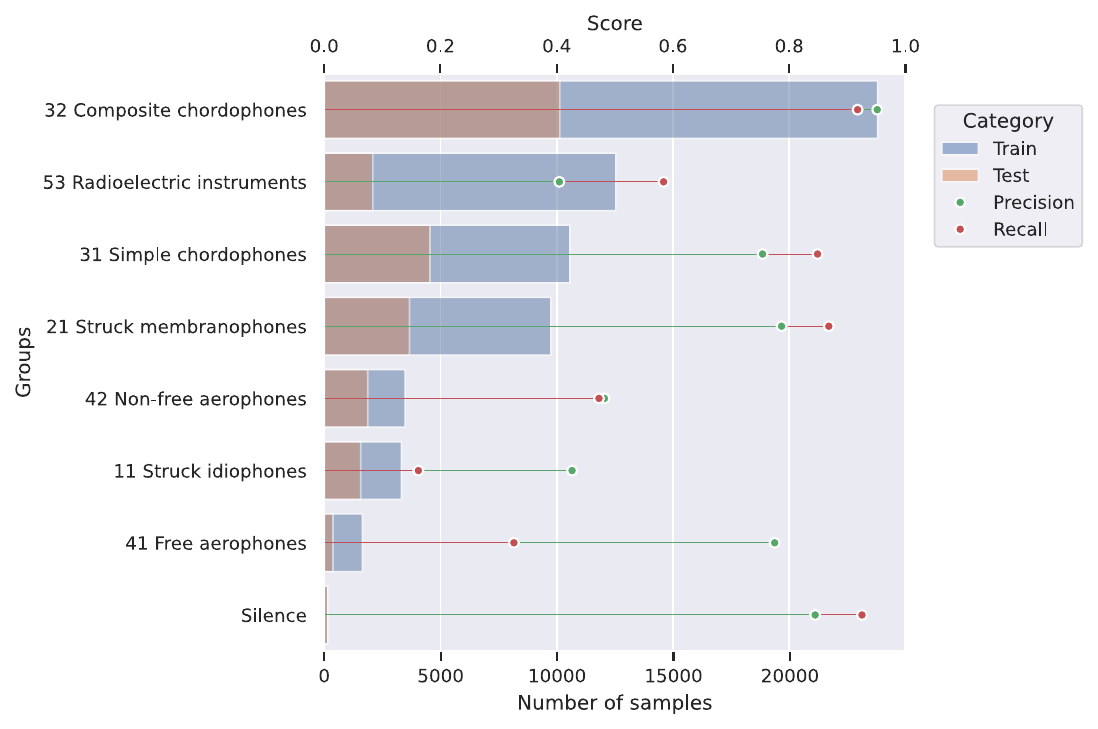}}
  \caption{\label{fig:loss_group}{\it Precision and recall per group, and as functions of the number of training and test samples for group-specialized models.}}
\end{figure}

This performance disparity is greatly lessened at a group level, as can be seen in \cref{fig:loss_group}. We can, however, notice the specific case of struck idiophones, which shows a much lower recall of $16\%$. Looking closer, we notice that this group is often misclassified as the \textit{Struck membranophones} group. That is not very surprising, given the considerable overlap between some of the instruments within each group. For instance, a gong or cymbals will be classified as \textit{Struck idiophones}, but any other auxiliary percussion will be considered a membranophone by default.
A non-negligible amount of \textit{Struck membranophones} instruments are also misclassified as \textit{Radioelectric instruments}: this is likely due to the presence of the drum machine in the latter group.
This shows the limitation of the chosen Hornbostel-Sachs class tree, which is very flexible in both depth and height but also can be prone to separating similarly-sounding instruments into very different groups.

\begin{figure}[!h]
  \centerline{\includegraphics[width=.95\linewidth]{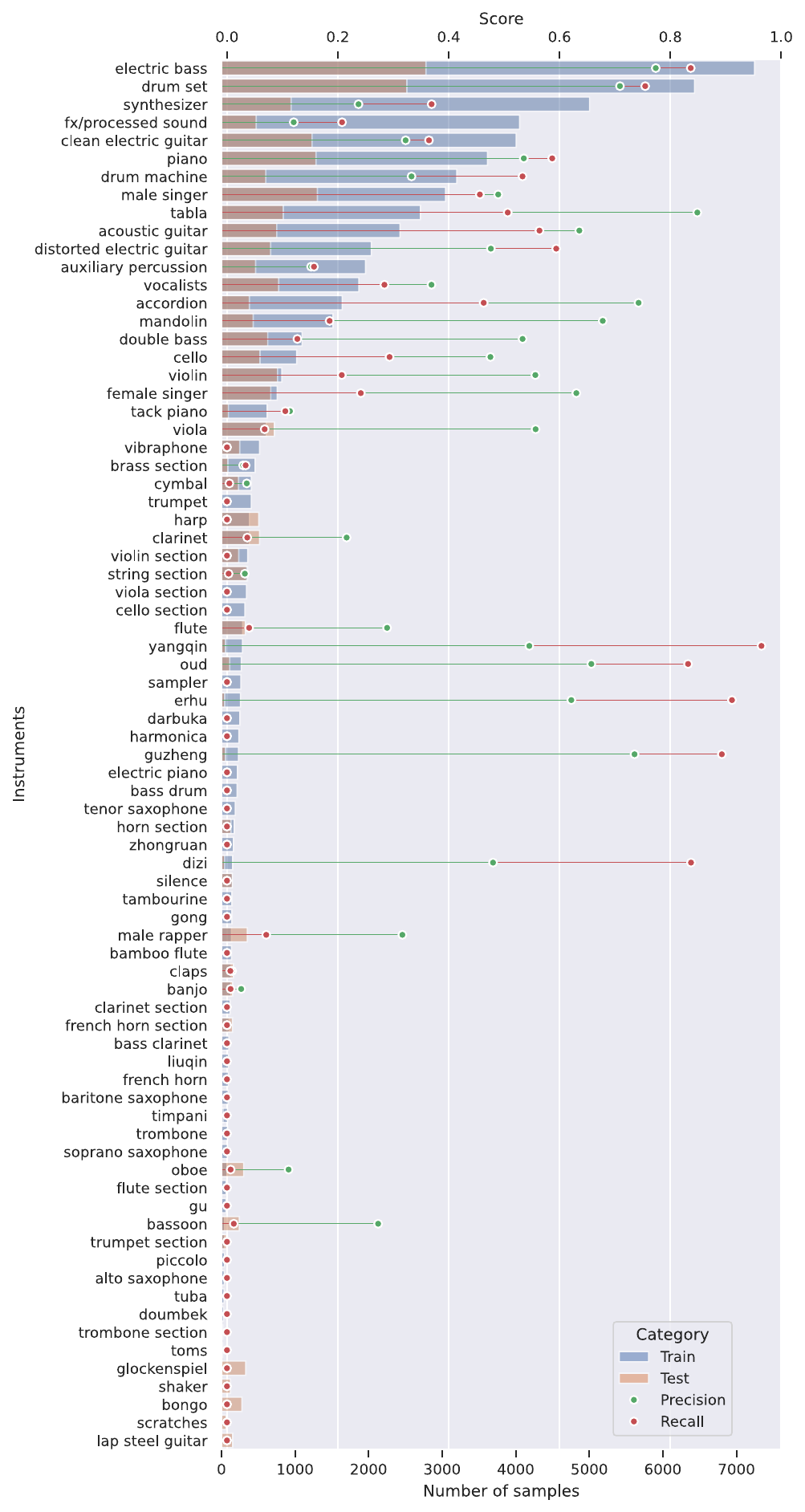}}
  \caption{\label{fig:loss}{\it Precision and recall per instrument, and as functions of the number of training and test samples for group-specialized models.}}
\end{figure}

We observed that our models are generally conservative, with recall scores notably lower than precision, especially at the instrument level. As depicted in \cref{fig:loss}, the model demonstrates reasonable performance for only about fifteen of the most common instruments, with performance sharply declining to almost zero precision and recall rates for most of the remaining dataset. Exceptions exist, with instruments such as the yangqin, the erhu, and the dizi showing some of the best performances. Given that these instruments all belong to traditional Chinese music, we can assume that the model has,  to a degree, learned to recognize this distinctive genre and its associated instruments.

\begin{figure}[!ht]
  \centerline{\includegraphics[width=.95\linewidth]{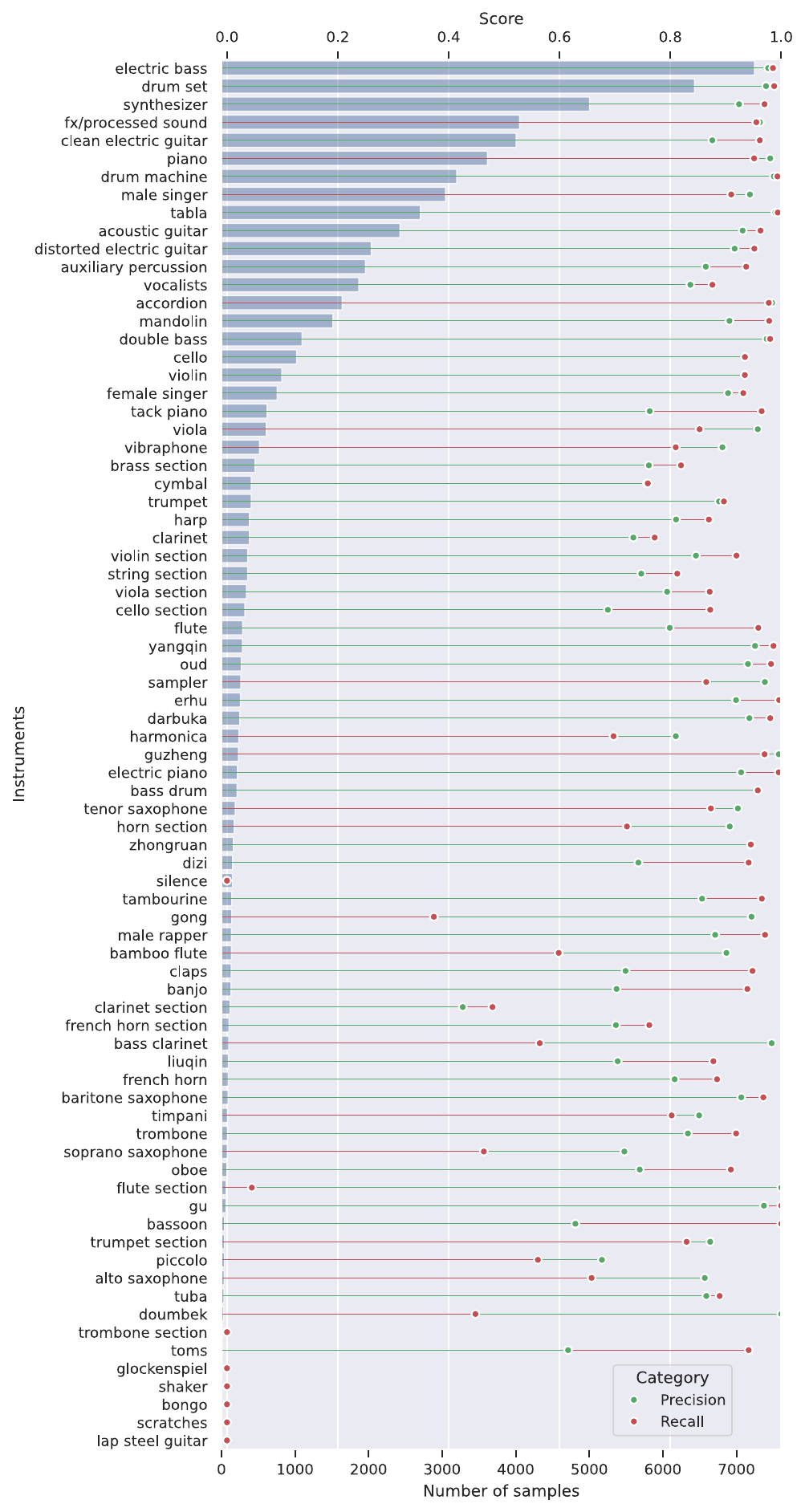}}
  \caption{\label{fig:loss_train}{\it Precision and recall for group-specialized models on the training data.}}
\end{figure}

\begin{figure}[!ht]
  \centerline{\includegraphics[width=\linewidth]{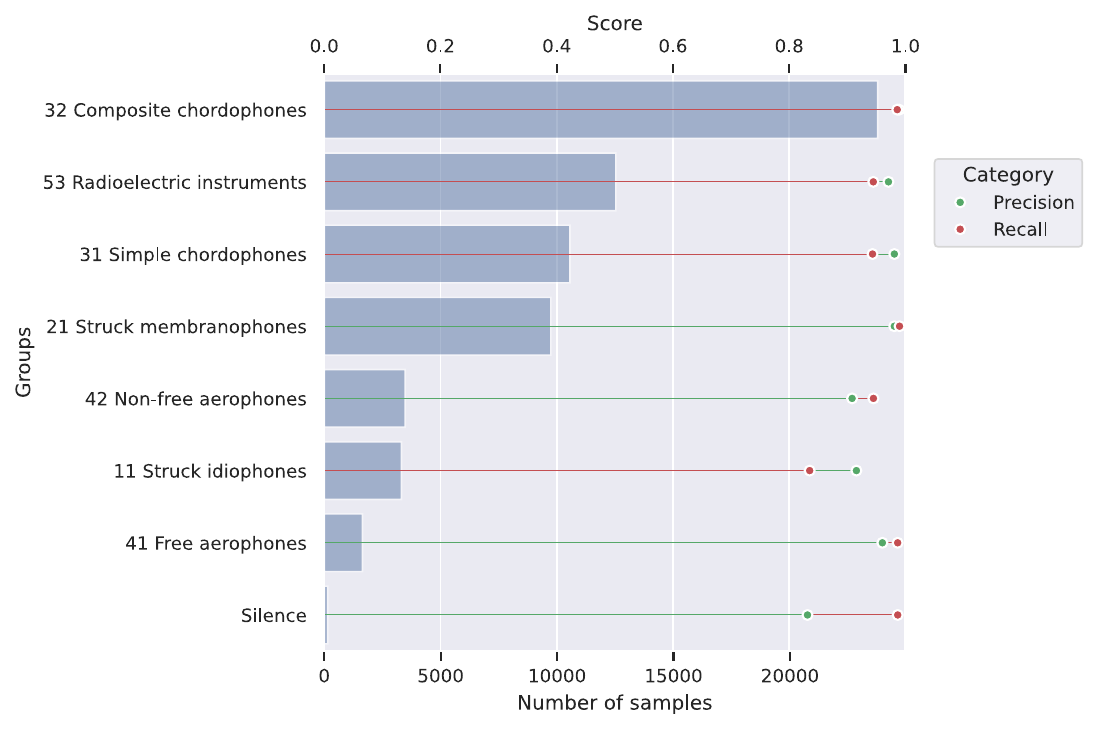}}
  \caption{\label{fig:loss_train_group}{\it Precision and recall on training data for group-specialized models.}}
\end{figure}

Furthermore, we are also able to confirm that the error of the model is caused by generalization issues. The performance of the model on the training set is excellent, as can be seen in \cref{fig:loss_train} and \cref{fig:loss_train_group}. Initial experiments with a validation set also allowed us to check that the model did not overfit the training data.

An important complicating element in instrument prediction lies in instrument co-occurrence. Let us define $C\in\mathbb{R}^{|\mathcal{I}| \times |\mathcal{I}|}$  where $C(i, j)$ represents the total instances of both the $i$th and $j$th instrument appearing together in a training set excerpt \footnote{\textcolor{black}{$|X|$ is the cardinal number of set $X$}}. This co-occurrence matrix is subsequently normalized within the range of [0, 1] utilizing the methodology outlined in \cite{huang2021multilabel}:

\begin{equation}
  \label{eq:norm}
  C'(i,j) =
  \begin{cases}
    0                                                                                   & \text{if } i = j  \\
    \dfrac{{C(i,j) - \min \, C(\cdot, j)}}{{\max \, C(\cdot, j) - \min \, C(\cdot, j)}} & \text{otherwise}.
  \end{cases}
\end{equation}
Because of the chosen normalization, the matrix is not symmetric and should be read "row-wise."
\begin{figure}[!ht]
  \centerline{\includegraphics[width=0.9\linewidth]{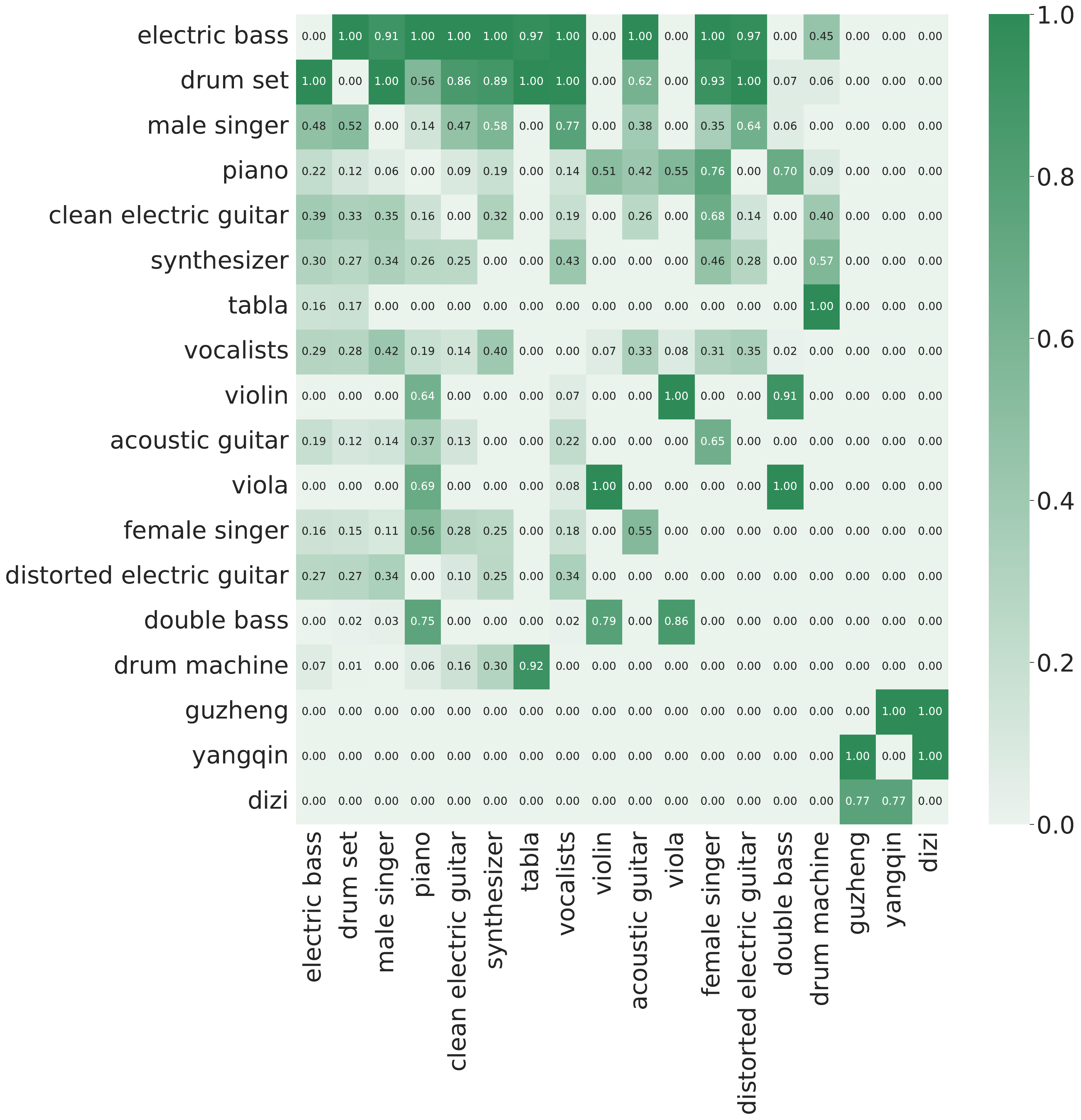}}
  \caption{\label{fig:cooc}{\it Excerpt of the normalized co-occurrence of instrument labels in the training data.}}
\end{figure}

\begin{figure}[!ht]
  \centerline{\includegraphics[width=0.9\linewidth]{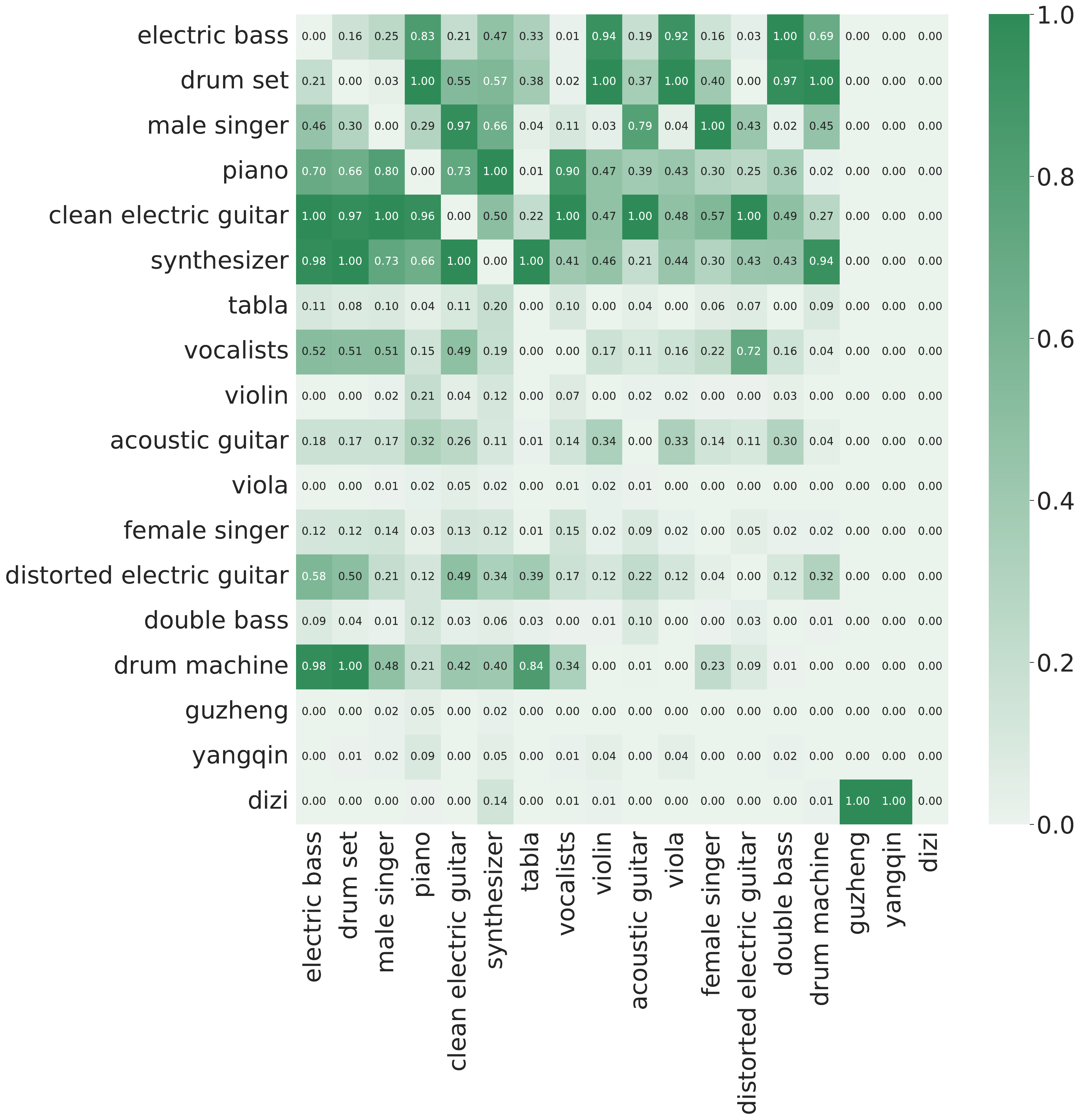}}
  \caption{\label{fig:fp}{\it False positive co-occurrence.
    }}
\end{figure}

\begin{figure}[!ht]
  \centerline{\includegraphics[width=0.9\linewidth]{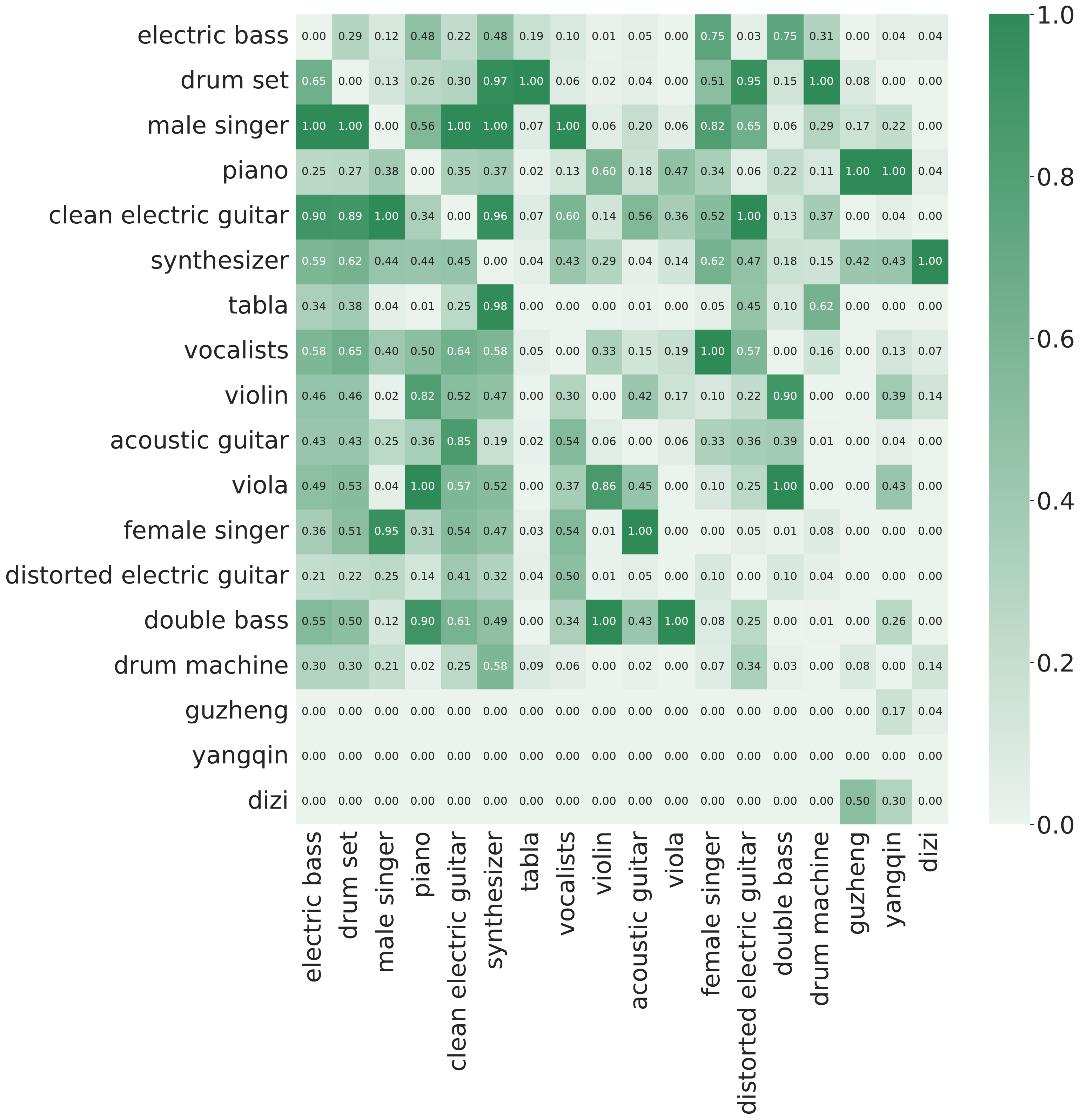}}
  \caption{\label{fig:fn}{\it False negative co-occurrence.
    }}
\end{figure}

An excerpt from this matrix, in \cref{fig:cooc}, shows these strong relations between some instruments.  For instance, we can confirm the speculated strong co-occurrence rate between Chinese instruments or notice that the violin and viola are always simultaneously present in the training data. Furthermore, displayed in \cref{fig:fp} (resp. \cref{fig:fn}) is a co-occurrence matrix illustrating instances of ghost detection (resp. missed detection). Specifically, within \cref{fig:fp}, the entry at $(i, j)$ denotes the occurrences of instrument $j$ in an excerpt when the model incorrectly predicted a false positive for $i$. In \cref{fig:fn}, the element at $(i, j)$ signifies the occurrences of predicted instrument $j$ in an excerpt where $i$ was erroneously identified as a false negative. These outcomes are standardized using the same methodology described in~\cref{eq:norm}, and the results should be read "row-wise."

Interestingly, instrument co-occurrence does not invariably result in false predictions, as the outcomes appear to be instrument-specific. Notably, the model's proficiency in recognizing instruments varies significantly. For instance, the model does not seem to have learned to effectively recognize the dizi, and seems to be detecting the guzheng and the yangqin as a proxy instead.
Besides, the model encounters challenges in distinguishing between specific instrument categories. For example, it frequently confuses digital drum machines with drum sets and mixes the double bass with the electric bass. An interesting fact is the ghostly detection of a distorted electric guitar when singers, electric bassists, and drum set players are present. This result aligns with expectations due to the widespread use of these instruments in Western music.

\section{Conclusion}\label{sec:ccl}
This paper shows that the hierarchical approach proves highly beneficial in rare instrument recognition within complex datasets. While the F1-score at an instrumental level shows poor performance of $45$\%, the group-level score reaches up to $78$\%, allowing for much more reliable coarse-level instrument detection.

Looking ahead, there are a few areas that could be explored further.
It would be interesting to investigate how to assess the system's adaptability to new instruments, particularly within established groups, to gauge its flexibility across various musical contexts. This would also allow us to bridge the gap between hierarchical systems and few-shot learning approaches.
\textcolor{black}{The current system's performance could also be evaluated on different datasets.}
Future works should also explore alternative input features for the neural network, such as audio scattering~\cite{anden2011multiscale}, and consider different hierarchical systems more tailored to machine learning methodologies. \textcolor{black}{The chosen instrument hierarchy is likely to have a strong impact on results, and} exploring automatic hierarchical classification for instruments~\cite{peeters2003automatic} represents an intriguing avenue to improve detection. Alternative model architectures should also be explored, such as the promising Vision Transformer-based models.
From the performance point of view, the specialized models can be simplified, making them smaller and faster to run. Such a study would make the models more efficient, which is crucial in real-world applications.


\bibliographystyle{IEEEbib}

\end{document}